# Google Auto ML versus Apple Create ML for Histopathologic Cancer Diagnosis; Which Algorithms Are Better?

Andrew A. Borkowski, MD*[1,2], Catherine P. Wilson, MT[1], Steven A. Borkowski[1], L. Brannon Thomas, MD[1,2], Lauren A. Deland, RN[1], Stefanie J. Grewe, MD[2], Stephen M. Mastorides, MD[1,2]

[1]Pathology and Laboratory Medicine Service, James A. Haley VA Hospital, Tampa, Florida, USA.

[2]Department of Pathology and Cell Biology, University of South Florida, Morsani College of Medicine, Tampa, Florida, USA.

*aborkows@health.usf.edu
Pathology and Laboratory Medicine Service (673/113)
James A Haley VA Hospital
13000 Bruce B. Downs Blvd
Tampa, FL 33612
Tel: 813-972-7525

**Keywords:** Google Auto ML, Apple Create ML, lung cancer, colon cancer, machine learning, AI, pathology, cancer diagnosis, medicine

**ABSTRACT**

Artificial Intelligence is set to revolutionize multiple fields in the coming years. One subset of AI, machine learning, shows immense potential for application in a diverse set of medical specialties, including diagnostic pathology. In this study, we investigate the utility of the Apple Create ML and Google Cloud Auto ML, two machine learning platforms, in a variety of pathological scenarios involving lung and colon pathology. First, we evaluate the ability of the platforms to differentiate normal lung tissue from cancerous lung tissue. Also, the ability to accurately distinguish two subtypes of lung cancer (adenocarcinoma and squamous cell carcinoma) is examined and compared. Similarly, the ability of the two programs to differentiate colon adenocarcinoma from normal colon is assessed as is done with lung tissue. Also, cases of colon adenocarcinoma are evaluated for the presence or absence of a specific gene mutation known as KRAS. Finally, our last experiment examines the ability of the Apple and Google platforms to differentiate between adenocarcinomas of lung origin versus colon origin. In our trained models for lung and colon cancer diagnosis, both Apple and Google machine learning algorithms performed very well individually and with no statistically significant differences found between the two platforms. However, some critical factors set them apart. Apple Create ML can be used on local computers but is limited to an Apple ecosystem. Google Auto ML is not platform specific but runs only in Google Cloud with associated computational fees. In the end, both are excellent machine learning tools that have great potential in the field of diagnostic pathology, and which one to choose would depend on personal preference, programming experience, and available storage space.

# 1. INTRODUCTION

Artificial intelligence (AI), first described in 1956, describes a field of computer science in which machines are trained to learn from experience. The term was popularized by the 1956 Dartmouth College Summer Research Project on Artificial Intelligence[1]. The field of AI is rapidly growing and has the potential to affect many aspects of our lives. The recent announcement from the White House by President Trump to launch the American AI initiative, allocating resources and funding for AI development, demonstrates this[2]. This initiative has a goal to not only improve American's quality of life but also to maintain national security and economic prosperity. As further research and development are conducted under this initiative, it is anticipated that rapid breakthroughs in AI technology will have a dramatic impact in multiple areas of society.

Machine learning, a subset of AI, was defined in 1959 by Arthur Samuel and is achieved by employing mathematical models to compute sample data sets[3]. Originating from statistical linear models, neural networks were conceived to accomplish these tasks[4]. These pioneering scientific achievements led to recent developments of deep neural networks. These models are created to recognize patterns and achieve computational excellence within a matter of minutes, often far exceeding human ability[5]. They can lead to increased efficiency by improved turnaround time due to decreased computation time, and high precision and recall[6].

Machine learning has the potential for numerous applications in the health care field[7][8][9]. One promising application is in the field of anatomic pathology, where specimens surgically removed from the patient are examined microscopically by Pathologists to render a diagnosis. Machine learning allows representative images to be used to train the computer to recognize patterns from labeled photographs. Based on a set of images selected to represent a specific tissue or disease process, the computer has the potential to be trained to evaluate and recognize imported images from patients, and render a diagnosis[10]. Prior to modern machine learning models, users would have to import many thousands of training images to produce algorithms that could recognize patterns with high accuracy. Modern machine learning algorithms allow for a model known as transfer learning, such that far fewer images are required for training[11]–[13].

Two novel machine learning platforms available for public use are offered through Google[14] and Apple[15]. They offer a user-friendly interface with minimal experience in computer science. Google AutoML is a beta release that utilizes machine learning via Cloud services to store and retrieve data with ease. No coding knowledge is required. Apple's machine learning platform, Apple Create ML Module, provides computer-based machine learning, requiring only a few lines of code to be operational.

We have previously reported using machine learning via the Apple Create ML Module in detecting non-small cell lung cancers (adenocarcinomas and squamous cell carcinomas) and colon cancers with very high accuracy[16][17]. In the present study, we expand on these findings by comparing the Apple platform to Google's AutoML platform. Using limited training data, both programs are compared for precision and recall in differentiating a variety of surgical pathology diagnoses. The lungs and colon are among the most common sites for cancers in America and are the top two causes of cancer-related deaths[18]. The vast majority of colon cancers are a type known as adenocarcinoma. Lung cancers commonly are classified into several subtypes based on morphologic and other characteristics. Among these, adenocarcinoma and squamous cell carcinoma, both in a group known as non-small cell carcinoma, represent the two most common types of lung cancer.

In this study, we investigate the utility of the two machine learning platforms to a variety of pathologic scenarios involving lung and colon pathology. First, we evaluate the ability of the platforms to differentiate normal lung tissue from cancerous lung tissue. Also, the ability to accurately distinguish two subtypes of lung cancer (adenocarcinoma and squamous cell carcinoma) is examined and compared.

Similarly, the ability of the two programs to differentiate colon adenocarcinoma from the normal colon is assessed as is done with lung tissue. In addition, cases of colon adenocarcinoma are assessed for the presence or absence of a specific gene mutation known as KRAS. KRAS is a proto-oncogene found in a variety of cancers, including approximately 40% of colon adenocarcinomas[19]. For colon cancers, the presence or absence of the KRAS mutation has important implications for patients as it determines whether the tumor will respond to specific chemotherapy agents[20]. The presence of the KRAS gene is currently determined by complex molecular testing of tumor tissue[21]. However, we assess the potential of machine learning to determine if the mutation is present by computerized morphologic analysis alone.

Finally, our last experiment examines the ability of the Apple and Google platforms to differentiate between adenocarcinomas of lung origin versus colon origin. Such information may be useful in determining the site of origin in tumors that have spread to other parts of the body[22].

## 2. MATERIALS AND METHODS

### 2.1 Image acquisition

Fifty cases of lung squamous cell carcinoma, fifty cases of lung adenocarcinoma, and fifty cases of colon adenocarcinoma were retrieved from our molecular database. 25 colon adenocarcinoma cases were positive for KRAS mutation, while 25 cases were negative for KRAS mutation. 750 total images of lung tissue (250 benign lung tissue, 250 lung adenocarcinomas, and 250 lung squamous cell carcinomas) and 500 total images of colon tissue (250 benign colon tissue and 250 colon adenocarcinoma) were obtained using a Leica Microscope MC190 HD Camera (Leica Microsystems, Wetzlar, Germany) connected to an Olympus BX41 microscope (Olympus Corporation of the Americas, Center Valley, PA, USA) and the Leica Acquire 9072 software for Apple computers. All the images were captured at a resolution of 1024 x 768 pixels using a 60x dry objective. Lung tissue images were captured and saved on a 2012 Apple MacBook Pro computer, and colon images were captured and saved on a 2011 Apple iMac computer. Both computers were running macOS v10.13 (Apple Inc, Cupertino, CA, USA). The Leica Acquire 9072 software works on macOS v10.13 but not on the latest macOS v10.14.

## 2.2 Creating Image Classifier Models using Apple Create ML

Apple Create ML is a suite of machine learning products that use tools such as the Swift programming language and the macOS playground to create and train custom machine learning models on mac computers[15]. The suite contains many features including Image Classification to train a machine learning model to classify images, Natural Language Processing to classify natural language text, and Tabular Data to train models that deal with labeling information or estimating new quantities. We used Create ML Image Classification to create Image Classifier Models for our project.

For each model, we followed the following procedure. We open the Apple Xcode macOS Playground v10 on a 2018 Apple MacBook Pro (macOS v10.14, 2.3 GHz Intel Core I5, 8 GB 2133 MHz LPDDR3, Intel Iris Plus Graphics 655 1536 MB, 256 SSD) to create Image Classifier Model with following lines of swift code:

```swift
import CreateMLUI
let builder = MLImageClassifierBuilder()
builder.showInLiveView()
```

We then opened the assistant editor in Xcode and ran the code. The live view displayed the image classifier UI. We then dragged in the training folder for training and validating the model. Once training and validation were complete, we dragged in the testing folder to evaluate the model performance on the indicated locations in live view. All training, validation, and testing was done at the default setting.

## 2.3 Creating ML modules using Google Cloud AutoML Vision beta

Google cloud AutoML is a suite of machine learning products including AutoML Vision, AutoML Natural Language and AutoML Translation[14]. All Cloud Auto ML machine learning products are currently in beta version. We used Cloud AutoML Vision beta to create machine learning modules for our project.

As opposed to the Apple Create ML which can be run on a local Apple computer, the Google Cloud Auto ML must be run in the cloud. The first step is to create a Google Cloud account. Although Google Cloud is a paid service, the first 12 months are offered as a free service with a $300 credit.

For each machine model, we used the following procedure. The main home screen for Google Cloud Platform is the console. The option for AutoML is on the upper left-hand side. Choosing the "Vision" option, leads to the welcoming screen and presented with a button labeled "get started with Vision." This option leads to the AutoML Vision beta Datasets screen. To create a new dataset, one has to choose "New Dataset" button located to the right of "AutoML Vision beta" button that is located in the upper left-hand corner of the screen. One then names the data set and chooses how to import images. We chose to import images later in order to be able to take advantage of the graphical user interface. We left classification type box unchecked because each of our image classes had only a single label. Next, we created class labels and imported images for each of the classes to the cloud. The following step involved training a new model. This step took on average 8 to 9 minutes. Once completed we received an email notification and we could evaluate our trained model. The following information was available: average precision (area under the precision-recall tradeoff curve), overall precision and recall, score threshold slider, precision-recall graphs, and confusion matrix. Confusion matrix table displays how often trained model classifies each label correctly and how often incorrectly ("confused model").

### 2.4 Experiment 1

In this experiment, we compared two machine learning algorithms, Apple Create ML Image Classifier and Google Auto ML Vision beta, in their ability to detect and subclassify non-small cell lung cancer based on the histopathological images.

Apple Create ML Image Classifier:

We created three classes of images (250 images each) with the following labels: Lung_Normal (benign lung tissue), Lung_AdenoCA (lung adenocarcinoma), and Lung_SqCA (lung squamous cell carcinoma). 80% of the images were randomly assigned to the training folder and 20% to the testing folder. The training folder included three class labeled subfolders with training images (80% of total). The testing folder included three class labeled subfolders with testing images (20% of total). Apple program randomly assigned 5% of images from training folder for validation.

Google Auto ML Vision beta:

We created three classes of images (250 images each) with the following labels: Lung_Normal (benign lung tissue), Lung_AdenoCA (lung adenocarcinoma), and Lung_SqCA (lung squamous cell carcinoma). 250 images per class (750 total images) were uploaded to Google Cloud. Google AutoML Vision beta randomly assigned 80% for training, 10% for validation, and 10% for testing the module.

### 2.5 Experiment 2

In this experiment, we compared two machine learning algorithms, Apple Create ML Image Classifier and Google Auto ML Vision beta, in their ability to differentiate between normal lung tissue and non-small cell lung cancer histopathologic images with 50/50 mixture of lung adenocarcinoma and lung squamous cell carcinoma.

Apple Create ML Image Classifier:

We created two classes of images (250 images each) with the following labels: Lung Normal (benign lung tissue) and Lung NSCLC (non-small cell lung cancer). 80% of the images were randomly assigned to the training folder and 20% to the testing folder. The training folder included two class labeled subfolders with training images (80% of total). The testing folder included two class labeled subfolders with testing images (20% of total). Apple program randomly assigned 5% of images from training folder for validation.

Google Auto ML Vision beta:

We created two classes of images (250 images each) with the following labels: Lung_Normal (benign lung tissue) and Lung NSCLC (non-small cell lung cancer). 250 images per class (500 total images) were uploaded to Google Cloud. Google AutoML Vision beta randomly assigned 80% for training, 10% for validation, and 10% for testing the module.

## 2.6 Experiment 3

In this experiment, we compared two machine learning algorithms, Apple Create ML Image Classifier and Google Auto ML Vision beta, in their ability to differentiate between lung adenocarcinoma and lung squamous cell carcinoma histopathologic images.

Apple Create ML Image Classifier:

We created two classes of images (250 images each) with the following labels: Lung AdenoCA (adenocarcinoma) and Lung SqCA (squamous cell carcinoma). 80% of the images were randomly assigned to the training folder and 20% to the testing folder. The training folder included two class labeled subfolders with training images (80% of total). The testing folder included two class labeled subfolders with testing images (20% of total).

Google Auto ML Vision beta:

We created two classes of images (250 images each) with the following labels: AdenoCA (lung adenocarcinoma) and SqCA (lung squamous cell carcinoma). 250 images per class (500 total images) were uploaded to Google Cloud. Google AutoML Vision beta randomly assigned 80% for training, 10% for validation, and 10% for testing the module.

## 2.7 Experiment 4

In this experiment, we compared two machine learning algorithms, Apple Create ML Image Classifier and Google Auto ML Vision beta, in their ability to detect colon cancer histopathological images regardless of KRAS mutation status.

Apple Create ML Image Classifier:

We created two classes of images (250 images each) with the following labels: Colon_Normal (benign colon tissue) and Colon_AdenoCA (colon adenocarcinoma). 80% of the images were randomly assigned to the training folder and 20% to the testing folder. The training folder included two class labeled subfolders with training images (80% of total). The testing folder included two class labeled subfolders with testing images (20% of total). Apple program randomly assigned 5% of images from training folder for validation.

Google Auto ML Vision beta:

We created two classes of images (250 images each) with the following labels: Colon_Normal (benign colon tissue) and Colon_AdenoCA (colon adenocarcinoma). 250 images per class (500 total images) were uploaded to Google Cloud. Google AutoML Vision beta randomly assigned 80% for training, 10% for validation, and 10% for testing the module.

## 2.8 Experiment 5

In this experiment, we compared two machine learning algorithms, Apple Create ML Image Classifier and Google Auto ML Vision beta, in their ability to differentiate between KRAS mutation-positive colon adenocarcinoma and KRAS mutation-negative colon adenocarcinoma histopathologic images.

Apple Create ML Image Classifier:

We created two classes of images (125 images each) with the following labels: KRAS_POS_AdenoCA (colon adenocarcinoma cases with KRAS mutation) and KRAS_NEG_AdenoCA (colon adenocarcinoma cases without KRAS mutation). 80% of the images were randomly assigned to the training folder and 20% to the testing folder. The training folder included two class labeled subfolders with training images (80% of total). The testing folder included two class labeled subfolders with testing images (20% of total). Apple program randomly assigned 5% of images from training folder for validation.

Google Auto ML Vision beta:

We created two classes of images (125 images each) with the following labels: AdenoCA_KRAS_Neg (colon adenocarcinoma cases without KRAS mutations) and AdenoCA_KRAS_Pos (colon adenocarcinoma cases with KRAS mutations). 125 images per class (250 total images) were uploaded to Google Cloud. Google AutoML Vision beta randomly assigned 80% for training, 10% for validation, and 10% for testing the module.

## 2.9 Experiment 6

In this experiment, we compared two machine learning algorithms, Apple Create ML Image Classifier and Google Auto ML Vision beta, in their ability to differentiate between lung adenocarcinoma and colon adenocarcinoma histopathologic images.

Apple Create ML Image Classifier:

We created two classes of images (250 images each) with the following labels: Colon_AdenoCA (colon adenocarcinoma) and Lung_AdenoCA (lung adenocarcinoma). 80% of the images were randomly assigned to the training folder and 20% to the testing folder. The training folder included two class labeled subfolders with training images (80% of total). The testing folder included two class labeled subfolders with testing images (20% of total). Apple program randomly assigned 5% of images from training folder for validation.

Google Auto ML Vision beta:

We created two classes of images (250 images each) with the following labels: AdenoCA_Lung (lung adenocarcinoma) and AdenoCA_Colon (colon adenocarcinoma). 250 images per class (500 total images) were uploaded to Google Cloud. Google AutoML Vision beta randomly assigned 80% for training, 10% for validation, and 10% for testing the module.

## 3. RESULTS

Twelve machine learning models were created in six experiments using the Apple Create ML and the Google Auto ML beta. Results of models' performances in terms of recall and precision data are summarized in Table 1. To investigate recall and precision differences between the Apple and the Google machine learning algorithms, we performed two-tailed distribution, paired t-tests. No statistically significant differences were found (p-value of 0.52 for recall and 0.60 for precision).

Overall, all of the models performed well in distinguishing between normal and neoplastic tissue for both lung and colon cancers. In subclassifying non-small cell lung cancer into adenocarcinoma and squamous cell carcinoma, the models were shown to have high levels of precision and recall. The models were also successful in distinguishing between lung and colonic origin of adenocarcinoma (Figures 1-4). However, both Apple Create ML Image Classifier and Google Auto ML Vision beta had trouble discerning colon adenocarcinoma with KRAS mutations from adenocarcinoma without KRAS mutations.

| ML Model | Classes | Apple Recall | Google Recall | Apple Precision | Google Precision |
|---|---|---|---|---|---|
| | Lung Normal | 100.00% | 100.00% | 98.04% | 100.00% |
| | Lung AdenoCA | 86.00% | 85.70% | 97.73% | 81.80% |
| Model 1 | Lung SqCA | 98.00% | 89.90% | 89.09% | 90.60% |
| | Lung Normal | 100.00% | 100.00% | 98.04% | 100.00% |
| Model 2 | NSCLC | 98.00% | 100.00% | 100.00% | 100.00% |
| | Lung AdenoCA | 90.00% | 82.40% | 91.84% | 87.50% |
| Model 3 | Lung SqCA | 92.00% | 92.30% | 90.20% | 88.90% |
| | Colon Normal | 98.00% | 95.80% | 100.00% | 100.00% |
| Model 4 | Colon AdenoCA | 100.00% | 100.00% | 98.04% | 96.00% |
| | Colon AdenoCA KRAS+ | 72.00% | 88.20% | 69.23% | 71.40% |
| Model 5 | Colon AdenoCA KRAS- | 68.00% | 50.00% | 70.83% | 75.00% |
| | Lung AdenoCA | 96.00% | 91.30% | 96.00% | 100.00% |
| Model 6 | Colon AdenoCA | 96.00% | 100.00% | 96.00% | 93.90% |

Table 1. Results summary for the Apple Create ML and the Google Auto ML beta.

## 4. DISCUSSION

Image classifier models utilizing algorithms via machine learning hold a promising future to revolutionize the healthcare field. Machine learning products, such as those modules offered by Apple and Google, are easy to use and have a simple graphical user interface to allow individuals to train models to perform human-like tasks in real time. In our experiments, we compared multiple algorithms to determine their ability to differentiate and subclassify histopathologic images with high precision and recall. To compare and contrast these modules, we focused our experiments on tasks that highlighted common and characteristic practices of histopathologic diagnosis and neoplasm classification.

Through six different experiments crafted to train, validate, and test different machine learning products, we examined the performance characteristics and features currently available. Analysis of the results revealed high precision and recall values illustrating the models' ability to differentiate and detect benign lung tissue from lung squamous cell carcinoma and lung adenocarcinoma in ML model one, benign lung from non-small cell lung carcinoma in ML model two, and benign colon from colonic adenocarcinoma in ML model 4. In ML model three and six, both machine learning algorithms performed at a high level to differentiate lung squamous cell carcinoma from lung adenocarcinoma and lung adenocarcinoma from colonic adenocarcinoma, respectively. Of note, ML model 5 had the lowest precision and recall values across both algorithms demonstrating the models' limited utility in predicting molecular profiles such as KRAS mutations as tested here. This is not surprising as pathologists currently require complex molecular tests to detect KRAS mutations reliably in colon cancer.

Both modules require minimal programming experience and are easy to use. In our comparison we demonstrate critical distinguishing characteristics that differentiate the two products.

Apple CreateML image classifier is available for use on local Mac computers that use Xcode version 10 and macOS 10.14, with just three lines of code required to perform computations. While this product is limited to Apple products, it is free to use and images are stored on the computer hard drive. Of unique significance, on the Apple system platform images can be augmented to alter the appearance to enhance model training. For example, imported images can be cropped, rotated, blurred, and flipped, in order to optimize the model's training abilities to recognize test images and perform pattern recognition. This feature is not available on the Google module. Apple CreateML Image classifier's default training set consists of 75% of total imported images with 5% of the total images being randomly utilized as a validation set. The remaining 20% of images comprise the testing set. The module's computational analysis to train the model is achieved in approximately 2 minutes on average. The score threshold is set at 50% and cannot be manipulated for each image class as in Google AutoML Vision beta.

Google AutoML Vision beta is open to many platforms and stores images on the Google Cloud, but it is important to note that it requires computing fees. Google gives new users 12 months of service for free with $300 for spending prior to charges required for model training functions. On AutoML Vision beta, random 80% of the total images are used in the training set, 10% are used in the validation set, and 10% are used in the testing set. It is important to highlight the different percentages used in the default settings on the respective modules. The time to train the Google

AutoML Vision beta with default computational power is longer on average than Apple CreateML, with approximately 8 minutes required to train the machine learning module. It is, however, possible to choose more computational power for an additional fee and decrease module training time. The user will receive an email when the computer time begins and an email when the computing is completed. This allows users to perform other tasks and return when the training is complete, at which time the user is notified via email. The computation time is calculated by subtracting the time of the initial email from the final email.

Based on our calculations we determined there was no significant difference between the two machine learning algorithms tested at the default settings with recall and precision values obtained. These findings demonstrate the promise of using a machine learning algorithm to assist in the performance of human tasks and behaviors, specifically the diagnosis of histopathologic images. These results have numerous potential uses in clinical medicine. Machine learning algorithms have been successfully applied to diagnostic and prognostic endeavors in pathology[23]–[28], dermatology[29]–[31], ophthalmology[32], cardiology[33], and radiology[34]–[36]. Pathologists often employ additional tests, such as special staining of tissues or molecular tests, to assist with accurate classification of tumors. Machine learning platforms offer the potential of an additional tool for Pathologists to utilize along with human microscopic interpretation[37][38]. In addition, many countries have marked physician shortages, especially in fields of specialized training such as pathology[39]–[41]. These models could readily assist physicians in these countries to treat patients with more specific diagnoses provided in a timely manner[42]. Finally, while we have explored the application of these platforms in common cancer scenarios, there is great potential to use similar techniques in the detection of other conditions. These include the potential for classification and risk assessment of precancerous lesions, infectious processes in tissue (e.g., detection of tuberculosis or malaria)[24][43], inflammatory conditions (e.g., arthritis subtypes, gout,)[44], blood disorders (e.g., abnormal blood cell morphology)[45] and many others. The potential of these technologies appears to be limited only by the imagination of the user[46].

Regarding the limited effectiveness in determining the presence or absence of KRAS mutations in colon adenocarcinoma, it is mentioned that currently, pathologists rely on complex molecular tests to detect the mutations at the DNA level[21]. It is possible that the use of more extensive training data sets may improve recall and precision in cases such as these and warrants further study. Our experiments were limited to the stipulations placed by the free trial software agreements; no costs were expended to utilize the algorithms, though an Apple computer was required.

## 5. CONCLUSIONS

In summary, Google Cloud Auto ML and Apple Create ML are democratizing machine learning. In our trained models for lung and colon cancer diagnosis, they both performed very well individually, and no statistically significant differences were found between the two platforms. However, there are some critical factors that set them apart. Apple Create ML can be used on local computers but is limited to an Apple ecosystem. Google Auto ML is not platform specific but runs only in Google Cloud with associated computational fees. In the end, both are excellent machine learning tools that have great potential in the field of diagnostic pathology, and which one to choose would depend on personal preference, programming experience, and available storage space.

**ACKNOWLEDGMENTS**

The authors would like to thank Paul Borkowski for his constructive criticism and proofreading of this manuscript.

**FUNDING**

This material is the result of work supported with resources and the use of facilities at the James A. Haley VA Hospital.

**REFERENCES:**

[1] J. Moor, "The Dartmouth College Artificial Intelligence Conference: The Next Fifty Years," *AI Mag.*, vol. 27, no. 4, pp. 87–87, Dec. 2006.
[2] "Accelerating America's Leadership in Artificial Intelligence." [Online]. Available: https://www.whitehouse.gov/articles/accelerating-americas-leadership-in-artificial-intelligence/. [Accessed: 12-Mar-2019].
[3] A. L. Samuel, "Some Studies in Machine Learning Using the Game of Checkers," *IBM J. Res. Dev.*, vol. 3, no. 3, pp. 210–229, Jul. 1959.
[4] W. S. Sarle, "Neural networks and statistical models," in *Proceedings of the nineteenth annual SAS users group international conference*, pp. 1538–1550, 1994.
[5] J. Schmidhuber, "Deep learning in neural networks: An overview," *Neural Networks*, vol. 61, pp. 85–117, Jan. 2015.
[6] Y. Lecun, Y. Bengio, and G. Hinton, "Deep learning," *Nature*, vol. 521, no. 7553. pp. 436–444, 2015.
[7] Y. Wang *et al.*, "Artificial intelligence in healthcare: past, present and future," *Stroke Vasc. Neurol.*, vol. 2, no. 4, pp. 230–243, 2017.
[8] B. J. Erickson, P. Korfiatis, Z. Akkus, and T. L. Kline, "Machine Learning for Medical Imaging," *RadioGraphics*, vol. 37, no. 2, pp. 505–515, 2017.
[9] R. C. Deo, "Machine Learning in Medicine HHS Public Access," *Circulation*, vol. 132, pp. 1920–1930, 2015.
[10] A. Janowczyk and A. Madabhushi, "Deep learning for digital pathology image analysis: A comprehensive tutorial with selected use cases," *J. Pathol. Inform.*, vol. 7, no. 1, p. 29, 2016.
[11] M. Oquab, L. Bottou, I. Laptev, and J. Sivic, "Learning and Transferring Mid-Level Image Representations using Convolutional Neural Networks." pp. 1717–1724, 2014.
[12] H. C. Shin *et al.*, "Deep Convolutional Neural Networks for Computer-Aided Detection: CNN Architectures, Dataset Characteristics and Transfer Learning," *IEEE Trans. Med. Imaging*, vol. 35, no. 5, pp. 1285–1298, 2016.
[13] N. Tajbakhsh *et al.*, "Convolutional Neural Networks for Medical Image Analysis: Full Training or Fine Tuning?," *IEEE Trans. Med. Imaging*, vol. 35, no. 5, pp. 1299–1312, 2016.
[14] "Cloud AutoML - Custom Machine Learning Models | AutoML | Google Cloud."

[Online]. Available: https://cloud.google.com/automl/. [Accessed: 12-Mar-2019].

[15] "Create ML | Apple Developer Documentation." [Online]. Available: https://developer.apple.com/documentation/createml. [Accessed: 12-Mar-2019].

[16] A. A. Borkowski, C. P. Wilson, S. A. Borkowski, L. A. Deland, and S. M. Mastorides, "Using Apple Machine Learning Algorithms to Detect and Subclassify Non-Small Cell Lung Cancer," **arXiv:1808.08230** [q-bio.QM] Aug. 2018.

[17] A. A. Borkowski, C. P. Wilson, S. A. Borkowski, L. B. Thomas, L. A. Deland, and S. M. Mastorides, "Apple Machine Learning Algorithms Successfully Detect Colon Cancer but Fail to Predict KRAS Mutation Status," **arXiv:1812.04660** [q-bio.QM] Dec. 2018.

[18] "American Cancer Society. Cancer Statistics Center." [Online]. Available: https://cancerstatisticscenter.cancer.org. [Accessed: 01-Nov-2018].

[19] T. Armaghany, J. D. Wilson, Q. Chu, and G. Mills, "Genetic alterations in colorectal cancer.," *Gastrointest. Cancer Res.*, vol. 5, no. 1, pp. 19–27, Jan. 2012.

[20] D. O. Herzig and V. L. Tsikitis, "Molecular markers for colon diagnosis, prognosis and targeted therapy," *J. Surg. Oncol.*, vol. 111, no. 1, pp. 96–102, Jan. 2015.

[21] W. Ma, S. Brodie, S. Agersborg, V. A. Funari, and M. Albitar, "Significant Improvement in Detecting BRAF, KRAS, and EGFR Mutations Using Next-Generation Sequencing as Compared with FDA-Cleared Kits," *Mol. Diagnosis Ther.*, vol. 21, no. 5, pp. 571–579, 2017.

[22] F. A. Greco, "Molecular Diagnosis of the Tissue of Origin in Cancer of Unknown Primary Site: Useful in Patient Management," *Curr. Treat. Options Oncol.*, vol. 14, no. 4, pp. 634–642, Dec. 2013.

[23] B. E. Bejnordi *et al.*, "Diagnostic assessment of deep learning algorithms for detection of lymph node metastases in women with breast cancer," *JAMA - J. Am. Med. Assoc.*, vol. 318, no. 22, pp. 2199–2210, 2017.

[24] T. Li, L. Chen, A. Hou, X. Ba, K. Zhang, and Y. Xiong, "Automatic detection of mycobacterium tuberculosis using artificial intelligence," *J. Thorac. Dis.*, vol. 10, no. 3, pp. 1936–1940, 2018.

[25] A. Cruz-Roa *et al.*, "Accurate and reproducible invasive breast cancer detection in whole-slide images: A Deep Learning approach for quantifying tumor extent," *Sci. Rep.*, vol. 7, no. 1, p. 46450, Dec. 2017.

[26] N. Coudray *et al.*, "Classification and mutation prediction from non–small cell lung cancer histopathology images using deep learning," *Nat. Med.*, vol. 24, no. 10, pp. 1559–1567, Oct. 2018.

[27] M. G. Ertosun and D. L. Rubin, "Automated Grading of Gliomas using Deep Learning in Digital Pathology Images: A modular approach with ensemble of convolutional neural networks.," *AMIA ... Annu. Symp. proceedings. AMIA Symp.*, vol. 2015, pp. 1899–908, 2015.

[28] N. Wahab, A. Khan, and Y. S. Lee, "Two-phase deep convolutional neural network for reducing class skewness in histopathological images based breast cancer detection," *Comput. Biol. Med.*, vol. 85, pp. 86–97, 2017.

[29] A. Esteva *et al.*, "Dermatologist-level classification of skin cancer with deep neural networks," *Nature*, vol. 542, no. 7639, pp. 115–118, Feb. 2017.

[30] I. Park *et al.*, "Deep neural networks show an equivalent and often superior performance to dermatologists in onychomycosis diagnosis: Automatic construction of onychomycosis datasets by region-based convolutional deep neural network," *PLoS One*, vol. 13, no. 1, p.

e0191493, 2018.
[31] Y. Fujisawa *et al.*, "Deep-learning-based, computer-aided classifier developed with a small dataset of clinical images surpasses board-certified dermatologists in skin tumour diagnosis," *Br. J. Dermatol.*, vol. 180, no. 2, pp. 373–381, Jun. 2019.
[32] V. Gulshan *et al.*, "Development and Validation of a Deep Learning Algorithm for Detection of Diabetic Retinopathy in Retinal Fundus Photographs," *JAMA*, vol. 316, no. 22, p. 2402, Dec. 2016.
[33] S. F. Weng, J. Reps, J. Kai, J. M. Garibaldi, and N. Qureshi, "Can Machine-learning improve cardiovascular risk prediction using routine clinical data?," *PLoS One*, vol. 12, no. 4, 2017.
[34] J.-Z. Cheng *et al.*, "Computer-Aided Diagnosis with Deep Learning Architecture: Applications to Breast Lesions in US Images and Pulmonary Nodules in CT Scans," *Sci. Rep.*, vol. 6, no. 1, p. 24454, Jul. 2016.
[35] W. Yang *et al.*, "Searching for prostate cancer by fully automated magnetic resonance imaging classification: deep learning versus non-deep learning," *Sci. Rep.*, vol. 7, no. 1, 2017.
[36] P. Lakhani and B. Sundaram, "Deep Learning at Chest Radiography: Automated Classification of Pulmonary Tuberculosis by Using Convolutional Neural Networks," *Radiology*, vol. 284, no. 2, pp. 574–582, 2017.
[37] D. Bardou, K. Zhang, and S. M. Ahmad, "Classification of Breast Cancer Based on Histology Images Using Convolutional Neural Networks," *IEEE Access*, vol. 6, no. 6, pp. 24680–24693, 2018.
[38] F. Sheikhzadeh, R. K. Ward, D. Van Niekerk, and M. Guillaud, "Automatic labeling of molecular biomarkers of immunohistochemistry images using fully convolutional networks," *PLoS One*, vol. 13, no. 1, 2018.
[39] I. Benediktsson, H., Whitelaw, J., Roy, "Pathology services in developing countries: a challange," *Arch Pathol Lab Med*, vol 131, pp. 1636-1639, 2007.
[40] D. Graves, "The impact of the pathology workforce crisis on acute health care," *Aust. Heal. Rev.*, vol. 31, no. 5, p. 28, 2010.
[41] "NHS pathology shortages cause cancer diagnosis delays | GM." [Online]. Available: https://www.gmjournal.co.uk/nhs-pathology-shortages-are-causing-cancer-diagnosis-delays. [Accessed: 12-Mar-2019].
[42] L. M. Abbott and S. D. Smith, "Smartphone apps for skin cancer diagnosis: Implications for patients and practitioners," *Australas. J. Dermatol.*, vol. 59, no. 3, pp. 168–170, 2018.
[43] G. Thoma, S. Jaeger, M. Poostchi, K. Silamut, and R. J. Maude, "Image analysis and machine learning for detecting malaria," *Transl. Res.*, vol. 194, pp. 36–55, 2018.
[44] D. E. Orange *et al.*, "Identification of Three Rheumatoid Arthritis Disease Subtypes by Machine Learning Integration of Synovial Histologic Features and RNA Sequencing Data," *Arthritis Rheumatol.*, vol. 70, no. 5, pp. 690–701, 2018.
[45] J. Rodellar, S. Alférez, A. Acevedo, A. Molina, and A. Merino, "Image processing and machine learning in the morphological analysis of blood cells," *Int. J. Lab. Hematol.*, vol. 40, pp. 46–53, May 2018.
[46] G. Litjens *et al.*, "A survey on deep learning in medical image analysis," *Med. Image Anal.*, vol. 42, pp. 60–88, 2017.

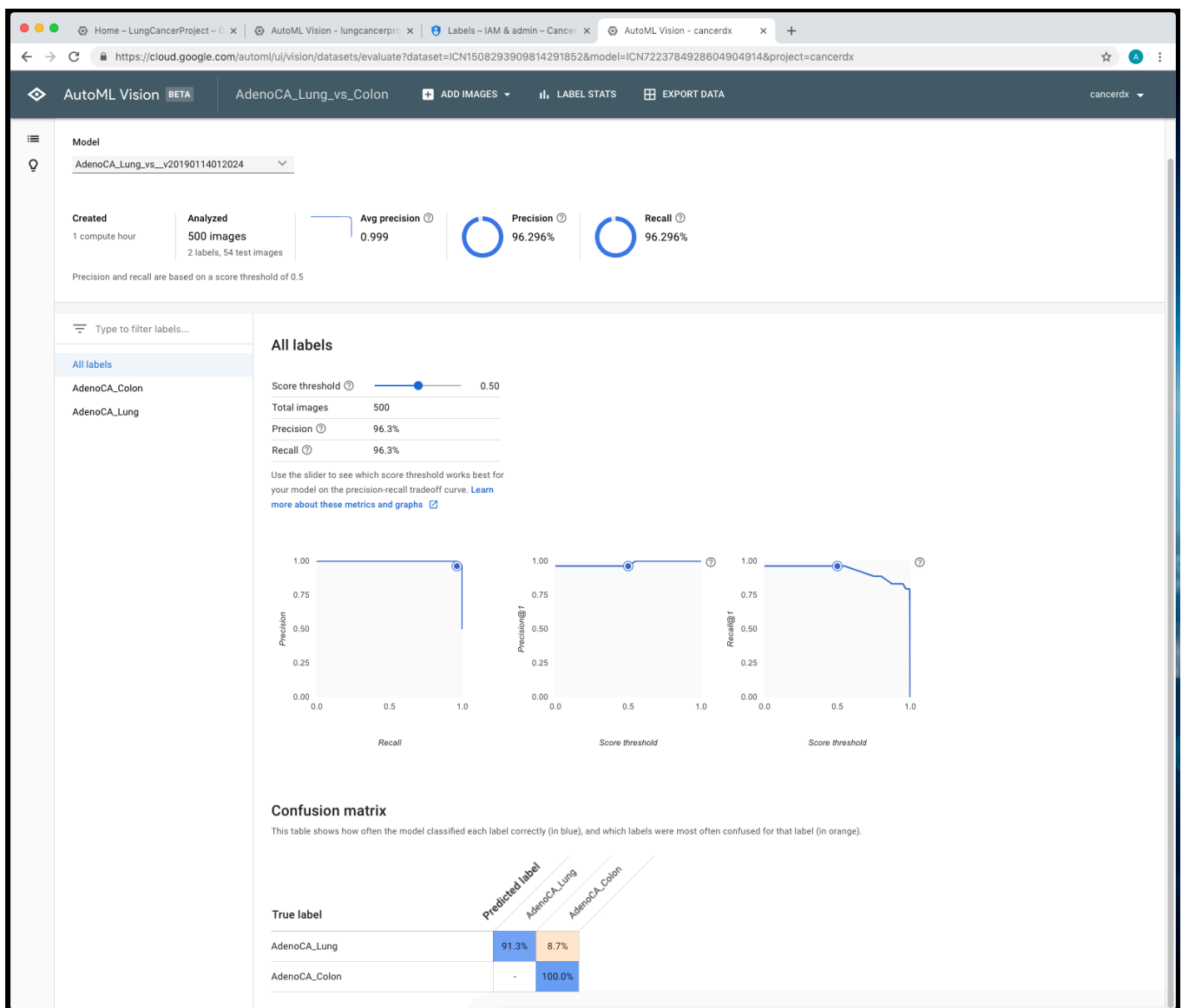

Figure 1. Screenshot of results using Google AutoML to differentiate lung adenocarcinoma from colon adenocarcinoma. Overall precision and recall data (Experiment 6).

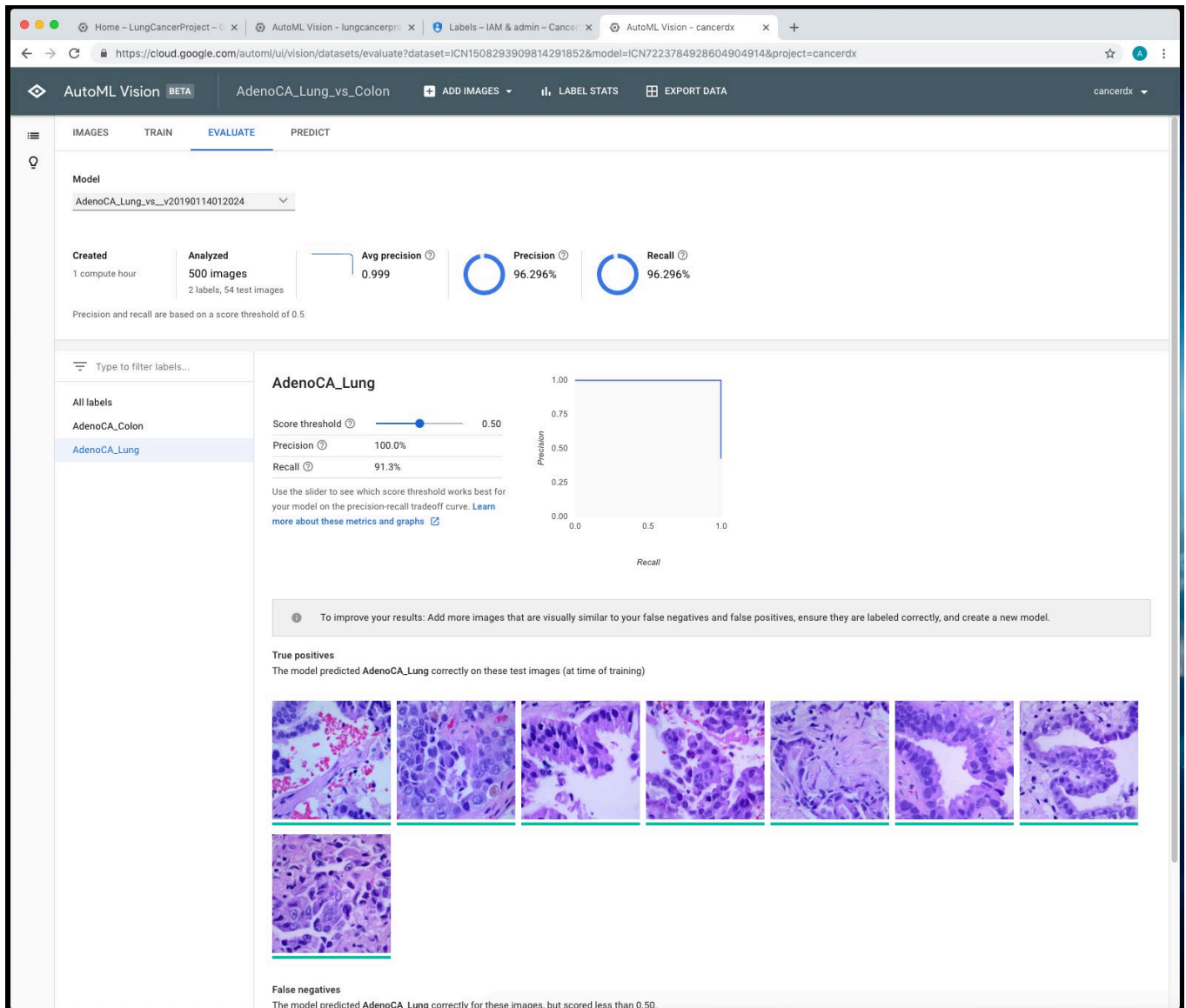

Figure 2. Screenshot of results using Google AutoML to differentiate lung adenocarcinoma from colon adenocarcinoma. Precision and recall data for lung adenocarcinoma label (Experiment 6).

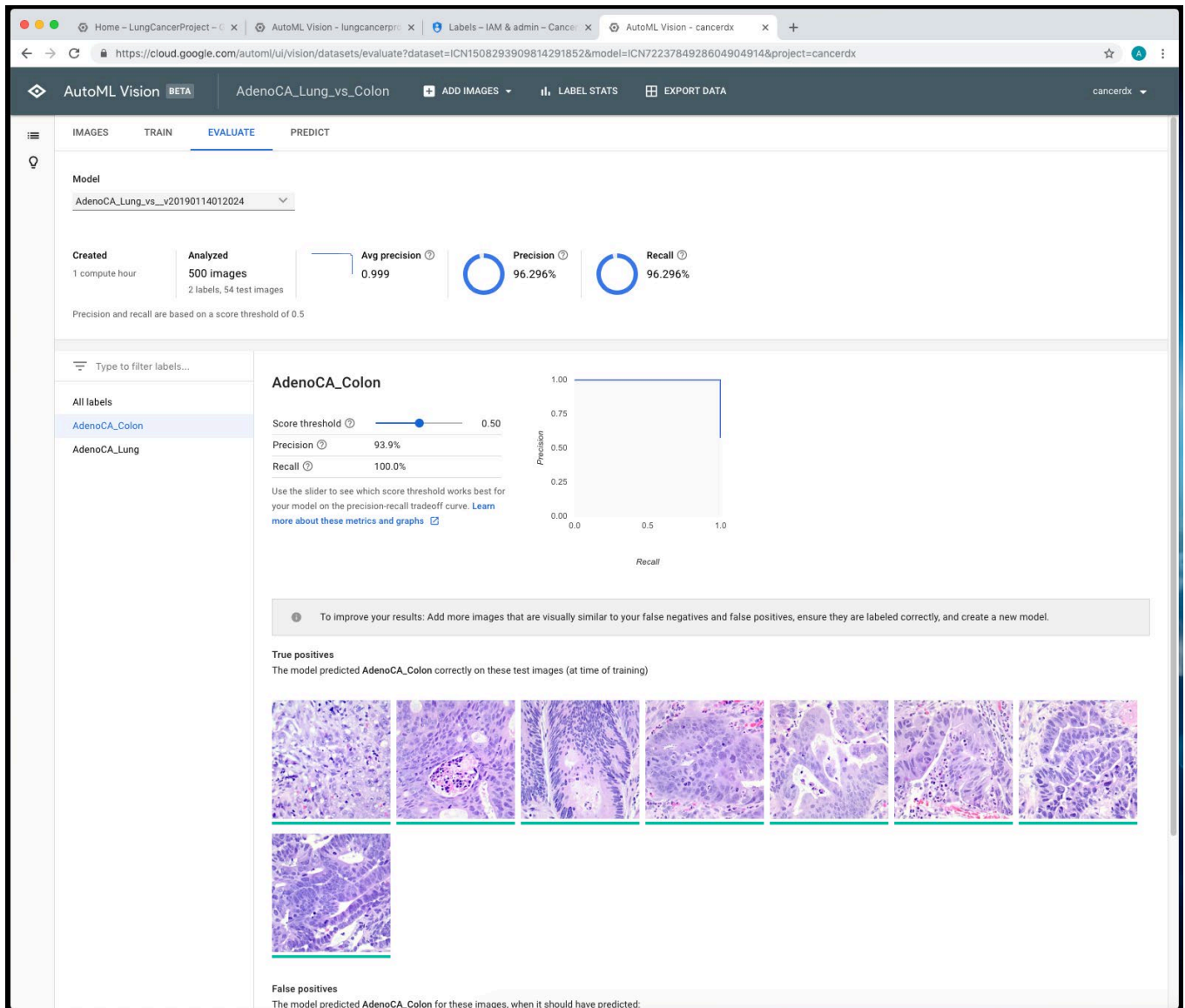

Figure 3. Screenshot of results using Google AutoML to differentiate lung adenocarcinoma from colon adenocarcinoma. Precision and recall data for colon adenocarcinoma label (Experiment 6).

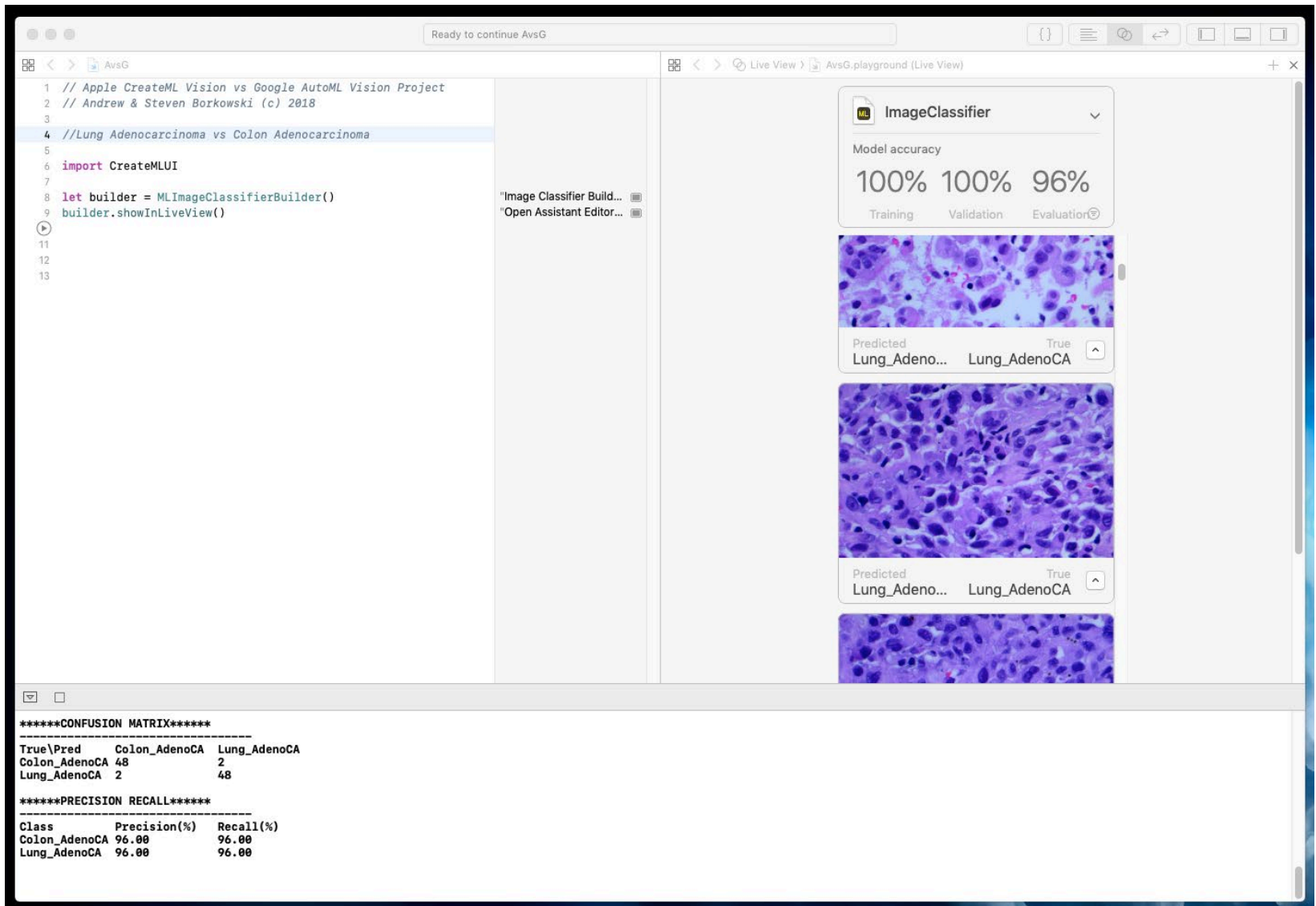

Figure 4. Screenshot of results using Apple Create ML demonstrating precision and recall differentiating lung adenocarcinomas from colon adenocarcinoma. Users can scroll through all images in the field on the right to review results for individual images (Experiment 6).